\documentclass[iop,revtex4,oneside]{emulateapj}
\usepackage{amsmath,amssymb}
\usepackage{multirow}
\usepackage{xcolor}
\usepackage{natbib}
\usepackage[colorlinks,citecolor=blue,linkcolor=blue,urlcolor=black,hyperfootnotes=true]{hyperref}
\usepackage{refcount}
\usepackage{array}
\usepackage{longtable}
\usepackage[figuresright]{rotating}
\usepackage{afterpage}
\usepackage{fancyhdr}
\usepackage{widetext}
\usepackage{amsmath}
\usepackage{amssymb}
\usepackage{ulem}
\usepackage{lipsum}
\usepackage[english]{babel}
\usepackage{array,kantlipsum}

\makeatletter

\afterpage{\global\setlength\@fpsep{8\p@ \@plus 2fil}}

\renewcommand\subsubsection{\@startsection{subsubsection}{3}{\z@}%
                                     {2.35ex\@plus -0.1ex \@minus -.0ex}%
                                     {0.4ex plus .1ex}%
                                     {\small\itshape \center}}
\patchcmd{\NAT@citex}
  {\@citea\NAT@hyper@{%
     \NAT@nmfmt{\NAT@nm}%
     \hyper@natlinkbreak{\NAT@aysep\NAT@spacechar}{\@citeb\@extra@b@citeb}%
     \NAT@date}}
  {\@citea\NAT@nmfmt{\NAT@nm}%
   \NAT@aysep\NAT@spacechar\NAT@hyper@{\NAT@date}}{}{}

\patchcmd{\NAT@citex}
  {\@citea\NAT@hyper@{%
     \NAT@nmfmt{\NAT@nm}%
     \hyper@natlinkbreak{\NAT@spacechar\NAT@@open\if*#1*\else#1\NAT@spacechar\fi}%
       {\@citeb\@extra@b@citeb}%
     \NAT@date}}
  {\@citea\NAT@nmfmt{\NAT@nm}%
   \NAT@spacechar\NAT@@open\if*#1*\else#1\NAT@spacechar\fi\NAT@hyper@{\NAT@date}}
  {}{}

\newcommand\@tmp{}
\newcommand\@refname{}
\newcommand\@refnames{}
\newcommand\@nextref{}

\newcommand\aref[1]{\@first@ref#1,@}
\def\@throw@dot#1.#2@{#1}% discard everything after the dot
\def\@set@refname#1{% set \@refname to autoefname(+s) using \getrefbykeydefault
    \edef\@tmp{\getrefbykeydefault{#1}{anchor}{}}%
    \def\@refname{\@nameuse{\expandafter\@throw@dot\@tmp.@autorefname}}%
    \def\@refnames{\@nameuse{\expandafter\@throw@dot\@tmp.@autorefname}s}%
}
\def\@first@ref#1,#2{%
    \@set@refname{#1}% set \@refname to autoref name
  \ifx#2@\let\@nextref\@gobble% only one ref
    \@refname~\ref{#1}% add singular autoefname and first reference
  \else%
    \@refnames~\ref{#1}% add plural autoefname and first reference
    \let\@nextref\@next@ref% push processing to \@next@ref
  \fi%
  \@nextref#2%
}
\def\@next@ref#1,#2{%
   \ifx#2@ and~\ref{#1}\let\@nextref\@gobble% at end: print and+\ref and stop
   \else, \ref{#1}% print  ,+\ref and continue
   \fi%
   \@nextref#2%
}

\makeatother

\begin{document}               % plus the \end{document} command at the end.

\makeatother

\title{Supernovae SDSS Host Metallicity Catalog \\and Resulting Metallicity Distributions by SNe Type}
\newcommand\pagetitle{\uppercase{Supernovae SDSS Host Metallicity Catalog}}
\newcommand\LastPage{\pageref{LastPage}}

\author{J. F. Graham$^{\hyperref[jfg]{1}}$ \affil{Kavli Institute for Astronomy and Astrophysics at Peking University,\\ No. 5 Yiheyuan Road, Haidian District, Beijing, P. R. China}}
\affil{$^1$\label{jfg}Kavli Institute for Astronomy and Astrophysics at Peking University, No. 5 Yiheyuan Road, Haidian District, Beijing, P. R. China}

\journalinfo{}
\submitted{}

\defcitealias{KobulnickyKewley}{KK04}
\defcitealias{T04}{T04}
\defcitealias{Dopita2016}{D16}
\defcitealias{kd2002}{KD02}

\begin{abstract}
In support of ongoing projects we have compiled a database of supernovae with host metallicities determined from spectroscopy in the Sloan Digital Sky Survey (SDSS) which we are providing as a catalog for the general use of the community.  Here we provide metallicity values in the KK04 \citep{KobulnickyKewley}, T04 \citep{T04}, and D16 \citep{Dopita2016} diagnostics as well as a compilation of the relevant line fluxes to allow users to compute their own metallicities.  The sample presented here has approximately 1700 objects with a metallicity in at least one diagnostic, with about 1200 objects in \citetalias{KobulnickyKewley} of which over 300 are core-collapse SNe or about twice the number of core-collapse objects present in the compilation of \cite{Kelly} reflecting the recent considerable improvement in SNe detection capabilities.  We find some differences in the metallicity distributions between the subtypes of both Type II and Type Ia SNe and speculate that metallicity may play a role in if Type II SNe events happen as IIb \& IIn or as standard Type II.

\end{abstract}

\section{Introduction}

We now know of two types of transient events, Long-duration Gamma-Ray Bursts (LGRBs) and Super Luminous Supernovae (SLSNe), that show a strong preference towards occurring in low metallicity environments (\citealt{Stanek2007, Kewley2007, conference_proceedings, stats_paper} and \citealt{Lunnan2014, PerleySLSNe2016, ChenMetal2016} respectively).  Short delay transients, such as these and core collapse supernovae (SNe) in general, would be expected to occur proportionally to the star-formation rate (SFR) unless something in their formation process is impeded by the presence of heavy elements.  The recent work of \cite{Modjaz2019} finds that the low metallicity preference of LGRBs is also reflected in the general broad-lined Type Ic SNe population (i.e.\ those events without an associated LGRB).

While such strong metallicity preferences almost certainly require an intrinsically  metallicity dependent formation process for these events, small scale deviations between the metallicity distributions of different transients can also serve as useful probes of differences in other physical properties.  Using the methodology of \cite{stats_paper} and \cite{diff_rate_letter} is also possible to compare transient metallicity distributions with the star-formation weighted metallicity distribution of the local universe.

As transient type correlates very strongly with stellar mass, differences in the relative rate of different events can provide insight into differences in the Initial Mass Function (IMF).  SNe are ideal for this because of the large population of objects now available.  \cite{Schneider2018} claims that the 30 Doradus starbust in the LMC has more massive (15 to 200 $M_\sun$) stars than would be expected from the Galactic IMF.  We are currently looking into using the sample collect here to probe for IMF variability as a function of metallicity.  The simple idea is to observe the metallicity distribution at difference mass ranges and compare.  

\cite{Hakobyan2014} found the relative rate of Type Ibc to Type II SNe to be significantly higher in merging than undisturbed galaxies.  The physical mechanism driving this effect is not yet known, but almost certainly must be driven by some type of environmental or stellar population difference.  Differences in the core-collapse to Type Ia SNe ratio also provides insight into the star-formation history and with a large enough sample can be compared to galactic environmental and other properties.

Here (in Table \ref{SNe_Z_table}) we provide metallicities of all known SNe occurring in galaxies with Sloan Digital Sky Survey (SDSS) host spectroscopy in three commonly used emission line diagnostics: the R$_{23}$ KK04 diagnostic of \cite{KobulnickyKewley}, the T04 diagnostic of \cite{T04}, and the D16 diagnostic of \cite{Dopita2016}.  We describe these diagnostic in \aref{Metallicity_Determination}.  To increase the utility of our sample, we also provide (in Table \ref{SNe_line_table}) the line fluxes used for our metallicity calculations to allow the community to calculate metallicities in other diagnostics as needed.

The sample presented here contains over 1700 objects (with a metallicity in at least one diagnostic), with about 1200 objects in the \citetalias{KobulnickyKewley} scale.  Of these, over 300 are core-collapse SNe or about twice the number of core-collapse objects present in the compilation of \cite{Kelly} reflecting the recent considerable improvement in SNe detection capabilities.  We also include Type Ia SNe hosts of which we have over twice as many.

\section{Methods}

\subsection{SNe Sample and SDSS Crossmatching}

We begin with a sample of approximately thirty-four thousand objects from the Transient Name Server (\url{http://wis-tns.weizmann.ac.il}) comprising essentially all known SNe from the modern era.  We also incorporate the approximately eight thousand SNe indexed by the NASA/IPAC Extragalactic Database (NED -- \url{http://ned.ipac.caltech.edu}).  However we find these to be almost entirely redundant with the Transient Name Server sample.  The principle advantage of incorporating the NED data is that the associated host galaxy is given by name for most of the NED indexed SNe.  For these objects we then employ a second NED lookup to retrieve the position of the host galaxy.  This greatly aids the process of matching the SNe with an SDSS target.  It is also helpful that this population of NED named hosts are weighted towards brighter, bigger, closer and thus more easily confused host galaxies.  

We then compare the location of the SNe (or, if available, NED host galaxy) with the SDSS photometric catalogue.  By identifying the SDSS galaxy hosting the SNe in the (for our purposes) almost complete photometric catalogue and then checking if that galaxy has SDSS spectroscopy we greatly improve our ability to correctly associate SNe to SDSS spectroscopic hosts.  This effectively operates as a rejection algorithm as we exclude SNe that match better with an object in the photometric catalogue even if that object also matches well with a different object in the spectroscopic catalogue.  We use the SNe type information from the Transient Name Server.

\subsection{SDSS Spectroscopic Sample}
\label{MPA-JHU}

For determining SDSS metallicities we employ the spectroscopic line list from the Max-Planck-Institut f\"{u}r Astrophysik - John Hopkins University (MPA-JHU) emission line analysis. See \url{http://www.mpa-garching.mpg.de/SDSS/} for the data products, their descriptions, and a more detailed citations list.  In order to fully reproduce our sample the user should specify SPECTROTYPE galaxy, SUBCLASS starforming or starburst, and also require non-zero {\it ugriz} CMODEL values.  The resulting SDSS galaxy metallicity sample consists of approximately 137 thousand galaxies.  A principle advantage of this sample is its uniformity, the galaxies are selected for spectroscopy and observed in a way completely independent from the properties of the SNe which they host.

%\subsection{Sample selection}

\subsection{Metallicity Diagnostics \& Determination}\label{Metallicity_Determination}

We adopt emission line metallicity diagnostics as, in particular oxygen, emission lines are the strongest and most easily observed metallicity indicator for other galaxies.  Since core-collapse SNe are the result of recent star-formation the requirement of ongoing star-formation activity to create the HII regions and associated nebular emission features is not an issue.  Furthermore, because both core-collapse SNe progenitors and the OB association stars that power nebular emission are short lived the emission line abundances represent the current metallicity of the star-forming gas in the SNe host rather than for stellar metallicities where the abundances reflect the luminosity-weighted average of all (including older) stars.

Here we provide metallicity values in the KK04 \citep{KobulnickyKewley}, T04 \citep{T04}, and D16 \citep{Dopita2016} diagnostics.  Although we favor \citetalias{KobulnickyKewley} we provide the others as a service to the community.  A description of each diagnostic is provided below.

\subsubsection{R$_{23}$ KK04 diagnostic of \cite{KobulnickyKewley}}
\label{KK04}

First proposed by Bernard Pagel in 1979 \citep{Pagel1979, Pagel1980}, the R$_{23}$ method is one of the most commonly used emission line metallicity diagnostics using the ratio of oxygen to hydrogen line strengths.  Doubly ionized oxygen, [O~III], has strong lines at 4959 and 5007 {\AA} and singly ionized [O~II] has a particularly strong, often unresolved, doublet at 3727 {\AA}.\footnote{As these lines are considerably brighter than the faint [O~III] 4363 {\AA} auroral line, R$_{23}$ is used extensively at moderately high redshifts where the 4363 {\AA} line is not practically observable.}  The metallicity independent 4861 {\AA} H$\beta$ line is conveniently located between these features.  The ratio of the sum of the fluxes of these oxygen lines divided by the H$\beta$ flux gives the equation for R$_{23}$ used in the classical application of this diagnostic (Equation \ref{eqn}).

\begin{equation}
\label{eqn}
 R_{23} = {[O~II] + [O~III] \over H\beta} = {I_{3727} + I_{4959} + I_{5007} \over I_{4861}}
\end{equation}

%\text{\footrefp{ud}}
%\footnolabel{ud}{Unresolved doublet}
 
Even when summing the [O~II] and [O~III] line fluxes\footnote{The 4959 {\AA} [O~III] line is actually quantum mechanically fixed at $\frac{1}{3}$ the 5007 {\AA} lines flux (see \citealt{070714Bpaper}).  Thus we actually use an optimal simultaneous combination of the 4959 {\AA} and 5007 {\AA} lines fluxes given the provided errors.  When the 4959 {\AA} [O~III] line is not available is assumed to have $\frac{1}{3}$ the 5007 {\AA} [O~III] flux.} the relationship between the total flux and the metallicity is dependent on the ionization.  In the classical application of this diagnostic\footnote{The most commonalty used classical R$_{23}$ recipe and scale is that of \citealt{M91} (M91).}, the R$_{23}$ value would be calculated (via Equation \ref{eqn}) and then converted to a metallicity by following a ionization contour determined using the [O~III] to [O~II] line ratio.  However metallicity has an effect on the relation between the ionization and the [O~III] to [O~II] line ratio which the classical methodology doesn't take into account.  \cite{kd2002} (KD02) solves this issue by using iterative fitting to determine the metallicity and ionization parameter concurrently.  Here we adopt the \cite{KobulnickyKewley} (KK04) scale of the R$_{23}$ diagnostic, a slightly updated version of the basic \citetalias{kd2002} methodology.

Another problem with strong line oxygen diagnostics is that they are double valued over physically meaningful abundance ratios.  At low metallicity, an increase in metallicity will result in an increase in oxygen line flux.  However, emission from infrared fine-structure lines serves as a cooling mechanism and at high metallicity this effect becomes dominant causing the oxygen line flux to decrease with increasing metallicity.  This causes two metallicity values (one upper branch and one lower branch) to generate the same R$_{23}$ line ratio.  To resolve this degeneracy we adopt the typical approach of using the ratio of the 6584 {\AA} [N~II] line to H$\alpha$ as a supplemental metallicity diagnostic.  While the [N~II] / H$\alpha$ diagnostic is not nearly as precise as R$_{23}$ it is sufficient to select between the R$_{23}$ branches.  

A known limitation of R$_{23}$ is that as the upper and lower branches converge (at the peak of the oxygen line flux) the relationship between metallicity and oxygen line flux (i.e.\ R$_{23}$ value) becomes nearly flat and any errors on the line flux measurements will push the metallicity away from the branch intersection point.  When plotting a large set of objects, this typically results in a visible but artificial gap in the R$_{23}$ metallicities at about log(O/H)+12 $\sim$ 8.4 in the \citetalias{KobulnickyKewley} scale.  For constancy with our prior work, in the \citetalias{KobulnickyKewley} diagnostic, we use the exact same code as used to determine the metallicities given in \cite{stats_paper} and \cite{XS_paper}.

\subsubsection{T04 diagnostic of \cite{T04}}

The \citetalias{T04} metallicity diagnostic uses a Bayesian fitting approach allowing use of all available emission lines and allowing the diagnostic to be employed when lines otherwise critical to other metallicity diagnostics are missing.  Still as \citetalias{T04} weights the input lines by their S/N ratio and oxygen typically has the strongest metallicity dependent emission \citetalias{T04} is effectively an oxygen line diagnostic in most cases.  (As described in \ref{KK04}, oxygen line strength first increases then decreases with metallicity, and the \citetalias{T04} Bayesian fitting must thus also rely on other lines to break this degeneracy.  As [N~II] is typically the best line for this purpose, the Bayesian fitting relies on it in much the same way as it is used with R$_{23}$.)  Furthermore, as R$_{23}$ already uses the brightest available oxygen lines it is not surprising that the two methods can be cross calibrated rather well \citep{KewleyEllison}.  Still, \citetalias{T04} and \citetalias{KobulnickyKewley} use different photoionization models and other calibration assumptions and therefore are quite separate in scaling.  Also, as \citetalias{T04} can be used without the 3727 {\AA} [O~II] line, it is not constrained by the z $>$ 0.0209 limit of R$_{23}$.

Unlike the \citetalias{KobulnickyKewley} and \citetalias{Dopita2016} diagnostics the \citetalias{T04} metallicities are given in the MPA-JHU emission line analysis (see \aref{MPA-JHU}), and the \citetalias{T04} diagnostic requires the non public\cite{BruzualCharlot} photoionization codes, therefore for the \citetalias{T04} diagnostic we only quote the values given in the MPA-JHU data products.

\subsubsection{D16 diagnostic of \cite{Dopita2016}}

A recent alternative to strong line oxygen diagnostics is the nitrogen sulfur method of \cite{Dopita2016}.  Singly ionized nitrogen, [N~II], has a strongly metallicity dependent line at 6584 {\AA} and singly ionized sulfur, [S~II], has a doublet at 6717 \& 6731 {\AA}.  As noted in the \citetalias{KobulnickyKewley} discussion, \aref{KK04}, the ratio of [N~II] / H$\alpha$ is itself a crude metallicity diagnostic.  However, due to its strong dependence on the ionization parameter it gives only a gross estimate of abundance unless the ionization is known.  Perviously the ionization was determined via the [O~III] to [O~II] (i.e.\ 5007 vs. 3727 {\AA} line) ratio.  However when these lines are known it is favorable to employ the more accurate R$_{23}$ method, therefore the ionization parameter corrected [N~II] / H$\alpha$ diagnostic was of limited utility.

\begin{equation}
\label{D16_eqn}
\begin{split}
& y = log\left({[N~II] \over [S~II]}\right) + 0.264~log\left({[N~II] \over H\alpha}\right) \\
& y = log\left({I_{6584} \over I_{6717} + I_{6731}}\right) + 0.264~log\left(I_{6584} \over I_{6563}\right) \\
& 12+log(O/H) \approx 8.77 + y \\
& 12+log(O/H) = 8.77 + y + 0.45~(y + 0.3)^5
\end{split}
\end{equation}

\cite{Dopita2016} uses the [N~II]~/~[S~II] ratio as an additional term to correct the [N~II] / H$\alpha$ ratio (Equation~\ref{D16_eqn}).  This diagnostic has the noted observation advantages of requiring a narrower spectral range (6563 to 6731 {\AA}) than the strong line oxygen diagnostics (3727 to 5007 {\AA})\footnote{Even when not talking into account the requirement for [N~II]~/~H$\alpha$ observations to break the branch degeneracy.}, not being double values, being calculated in a simple equation not requiring iterative fitting, and being usable at z = 0 in the SDSS spectroscopy.\footnote{The requirement of the 3727 {\AA} line and the limits of the SDSS spectral wavelength coverage limit these observations to z = 0.0209 or higher.} The last of which is the reason that, of the different metallicity diagnostics used here, \citetalias{Dopita2016} has the largest number of objects.  

\section{Results}

The primary results of this work are the two long tables provided at the end.  Table \ref{SNe_Z_table} provides the host galaxy metallicity (in at least one of the three diagnostics) for 1847 SNe.  In Table \ref{SNe_line_table} (beginning on page \pageref{SNe_Z_table}) we provide the lines fluxes used to determine the metallicities given in Table \ref{SNe_Z_table}.  This allows the reader to compute their own metallicities using different diagnostics and scales without having to reproduce the lines flux list themselves.  

\newcolumntype{L}[1]{>{\centering\arraybackslash}m{#1}}
\begin{table}[t]
\begin{center}
\vspace{-0.2 cm}
%\begin{minipage}[H]{0.45\textwidth}
\defcitealias{KobulnickyKewley}{KK04}
\caption{\label{SNe_type_num_table} Supernovae with \citetalias{KobulnickyKewley} metallicities by type \vspace{-0.2 cm}}
%\end{minipage}
\begin{tabular}{m{3.3cm}L{1.7cm}L{1.7cm}}
\hline
\hline
\vspace{0.05 cm}
\multirow{ 2.3}{*}{SNe Type} & \multicolumn{2}{c}{Number of SNe} \\
\cline{2-3}
 & inclusive$^a$ & exclusive$^b$ \\
\hline
I &         893 &            6\\
II &         296 &          202\\
II-non-std &	94 &	\\
IIL &            \multicolumn{2}{c}{2}\\
IIP &           \multicolumn{2}{c}{44}\\
IIb &           \multicolumn{2}{c}{13}\\
IIn &           \multicolumn{2}{c}{35}\\
Ia &         820 &          783\\
Ia-pec &          37 &           17\\
Ia-CSM &            \multicolumn{2}{c}{1}\\
Ia-91T-like &           \multicolumn{2}{c}{13}\\
Ia-91bg-like &            \multicolumn{2}{c}{5}\\
Iax[0IIcx-like] &            \multicolumn{2}{c}{1}\\
Ib/c &         67  &            7\\
Ib &          25 &           23\\
Ibn &            \multicolumn{2}{c}{2}\\
Ic &          35 &           31\\
Ic-BL &            \multicolumn{2}{c}{4}\\
core-collapse &         363 &            \\
untyped &	\multicolumn{2}{c}{4}\\

\hline
\end{tabular}
\\
\begin{minipage}[H]{0.45\textwidth}
\vspace{0.1 cm}
\item $^a$Includes other listed subtypes\\
$^b$Excludes other listed subtypes
\end{minipage}
\end{center}
\end{table}

\begin{figure*}[t]
\begin{center}
\includegraphics[width=1\textwidth]{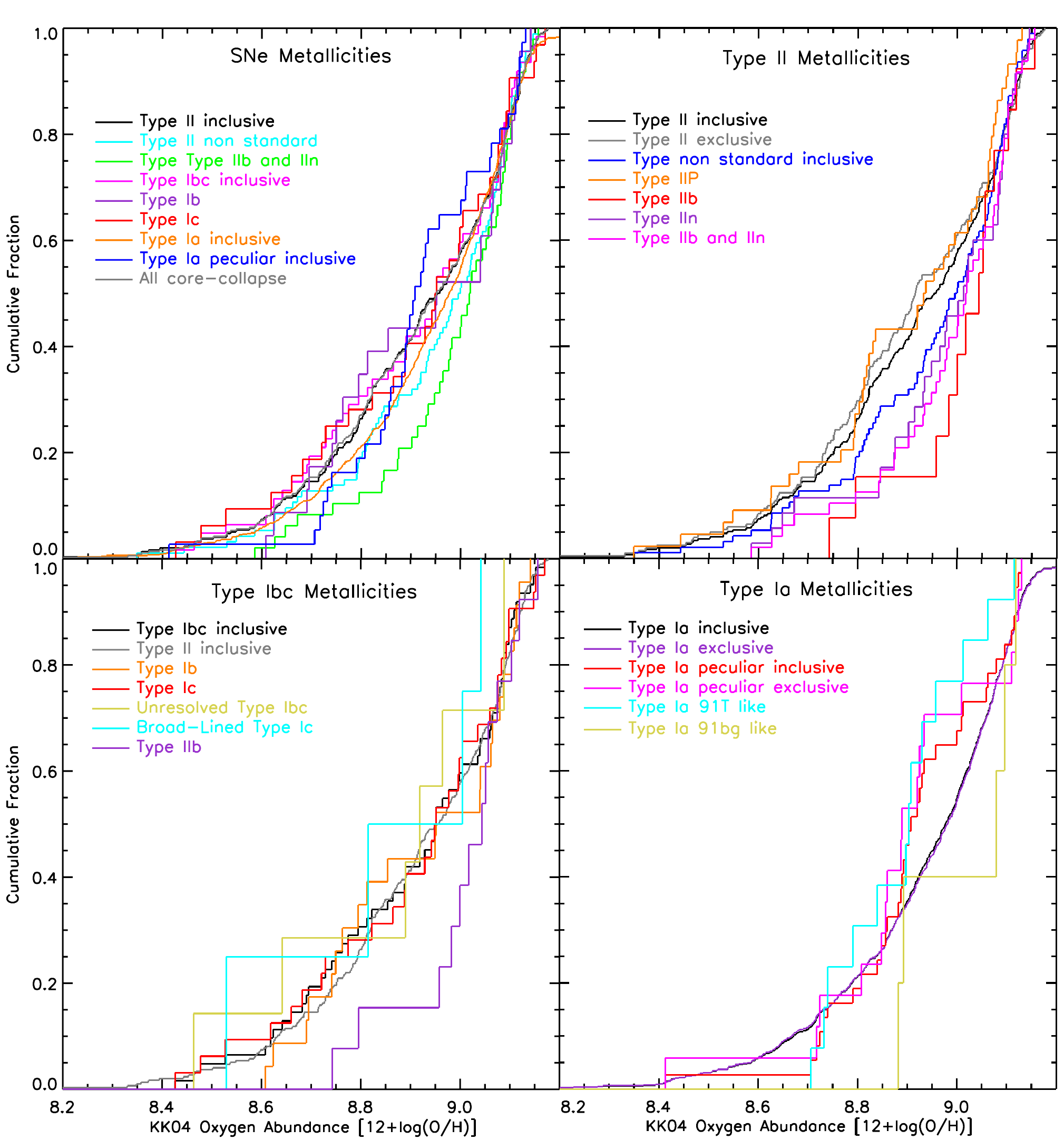}
\caption{\label{ZD} Cumulative distribution of \citetalias{KobulnickyKewley} SNe host metallicities of different SNe Types.  The top left pane compares the different basic SNe types and highlights the most discrepant populations.  The remaining panes each show the subtype populations individually.  KS probabilities comparing the metallicity distributions of the different populations are given in \aref{KS_table}.}

\end{center}
\end{figure*}

In \aref{SNe_type_num_table} we provide an accounting of the breakdown of SNe with \citetalias{KobulnickyKewley} metallicities by type.  Using this we can also update the \cite{form_rate_letter} fraction of ccSNe that are Type Ibc to $f_{Ibc/cc}$ = 18.4$\pm$2.4\% and the fraction of Type Ibc that are Type Ic to $f_{Ic/Ibc}$ = 41$\pm$10\%.  In \aref{ZD} we provide normalized cumulative distribution plots of the SNe host\citetalias{KobulnickyKewley} metallicity distributions of different SNe types.  We also compare these populations using KS statistics in \aref{KS_table}.

As a result of recent considerable improvements in SNe detection capabilities the number of SNe is increasing rapidly.  Throughout this project we have taken care to automate the process used and are investigating how to best make an updated version of this sample available on a frequent basis.

\section{Conclusions}

We find a number of clearly statically significant differences in the metallicity distributions of different SNe subtypes within the same Type however in general the different main core-collapse SNe Types agree reasonably well.  Notably while the Type IIP SNe track the exclusive Type II's (those Type II SNe without an additional subtype) the other subtypes (IIb and IIn show a presence for higher metallicities).  As we see no difference in the metallicity distribution between the Type IIb and IIn we have created a combined population to increase numbers.

Interestingly we find good agreement between the inclusive Type II population and the Type Ibc populations.  Both the exclusive and non-standard Type II populations have poorer agreement with the Ibc's but in opposing directions.  It is unclear if this is coincidence or if metallicity plays a role in whether a Type II SNe forms as a standard event or one of the Type II subtypes. 

As expected the Type Ia SNe do not follow the ccSNe metallicity distribution but  instead favor higher metallicities.  We also find that the peculiar and the 91T like Ia's favor lower metallicities than the standard Ia's.  The 91bg like events may favor higher metallicities than standard however more events are needed to know this conclusively. 

 In general the recent considerable improvement in SNe detection capabilities is resulting a considerable increase in SNe detected per year so perhaps we will have the answer to the environmental preferences of SNe subtypes (such as 91bg like events) soon.  Based on the results here it appears that either metallicity itself or a property that correlates with metallicity is causing a mild metallicity preferences in either the subtype of SNe explosion or in the rate of different subtypes.  While more data is needed (and should be available in the next few years), we speculate that metallicity may effect the type but not the rate of Type II SNe events.  I.e. they still happen, and happen as Type II's but the likelihood of their happening as a Type IIb or IIn may be affected by metallicity.

\acknowledgments

John Graham acknowledges support through the National Science Foundation of China (NSFC) under grant  11750110418.

\bibliographystyle{apj_links}
\bibliography{catalog_paper}

\renewcommand{\floatpagefraction}{1}
\newcolumntype{L}[1]{>{\centering\arraybackslash}m{#1}}
\begin{sidewaystable*}[p!]

\vspace{5 cm}%\hspace{0.1 cm}
\begin{minipage}[H]{0.9\textwidth}
{\normalsize\vspace{0.3 cm}\hspace{3.0 cm}\begin{NoHyper}\pageref{KS_table}\space/\space\LastPage\end{NoHyper}\hfill\hfill
\hspace{0.24 cm}\pagetitle
\hfill
GRAHAM \hspace{-0.0 cm}}
\vspace{0.2 cm}
\caption{\label{KS_table} KS probabilities for metallicity distributions of various SNe Types\vspace{-0.15 cm}}
\end{minipage}
\vspace{-0.3 cm}
%\begin{center}
%\tiny
\hspace{1 cm}\begin{tabular}{m{4.1cm}|L{1.2cm}L{1.5cm}L{1.5cm}L{1.5cm}L{1.2cm}L{1.5cm}L{1.2cm}L{1.2cm}L{1.2cm}L{1.2cm}L{1.2cm}L{1.2cm}L{1.2cm}}
\hline
\hline
 & Type Ia inc & Type Ia exc & Type Ia pec inc & Type Ia pec exc & Type Ia 91T like & Type Ia 91T like \& pec exc & Type Ia 91bg like & Type II inc & Type II exc & Type II non-std\\
\hline
Type Ia exc &     1.0000\\
Type Ia pec inc &     0.1041 &     0.0787\\
Type Ia pec exc &     0.1120 &     0.0925 &     0.9897\\
Type Ia 91T like &     0.1425 &     0.1224 &     0.9838 &     0.9857\\
Type Ia 91T like \& pec exc &     0.0180 &     0.0131 &     0.9996 &     1.0000 &     1.0000\\
Type Ia 91bg like &     0.4203 &     0.4231 &     0.4331 &     0.4183 &     0.1881 &     0.3011\\
Type II inc &     0.0369 &     0.0387 &     0.4308 &     0.3578 &     0.2800 &     0.0876 &     0.3457\\
Type II exc &     0.0091 &     0.0069 &     0.3694 &     0.7057 &     0.4848 &     0.2517 &     0.2447 &     0.7895\\
Type II non-std &     0.8199 &     0.8538 &     0.0614 &     0.0629 &     0.0872 &     0.0112 &     0.5691 &     0.1909 &     0.0213\\
Type IIP &     0.0999 &     0.1053 &     0.2724 &     0.6236 &     0.5894 &     0.4297 &     0.2594 &     0.6795 &     0.7384 &     0.3510\\
Type IIb &     0.1393 &     0.1619 &     0.0111 &     0.0130 &     0.0076 &     0.0024 &     0.5981 &     0.0779 &     0.0382 &     0.3631\\
Type IIn &     0.4427 &     0.4591 &     0.0508 &     0.0433 &     0.0629 &     0.0114 &     0.9516 &     0.0676 &     0.0270 &     0.5647\\
Type IIb \& IIn &     0.1490 &     0.1938 &     0.0084 &     0.0114 &     0.0184 &     0.0014 &     0.8463 &     0.0242 &     0.0028 &     0.5212\\
Type Ibc inc &     0.5113 &     0.5263 &     0.4765 &     0.3072 &     0.5025 &     0.1402 &     0.4481 &     0.9528 &     0.7592 &     0.4156\\
Type Ib &     0.5069 &     0.5061 &     0.5210 &     0.4072 &     0.2850 &     0.2240 &     0.3191 &     0.8582 &     0.7324 &     0.5548\\
Type Ic &     0.8775 &     0.8908 &     0.5610 &     0.3459 &     0.5207 &     0.2008 &     0.4522 &     0.9705 &     0.8080 &     0.7463\\
Unresolved Type Ibc &     0.7481 &     0.7098 &     0.7598 &     0.8848 &     0.7788 &     0.8025 &     0.8712 &     0.9048 &     0.9857 &     0.5702\\
Type Ic-bl &     0.8736 &     0.8751 &     0.8777 &     0.9463 &     0.9496 &     0.9753 &     0.4772 &     0.9887 &     0.9942 &     0.8766\\
All core-collapse &     0.0270 &     0.0295 &     0.4597 &     0.3238 &     0.2994 &     0.0969 &     0.3552 &     1.0000 &     0.6296 &     0.1979\\
All Type I &           1.0000 &     1.0000 &     0.1113 &     0.1175 &     0.1553 &     0.0194 &     0.4324 &     0.0561 &     0.0136 &     0.8155\\

\hline
\hline
 & Type IIP & Type IIb & Type IIn & Type IIb \& IIn & Type Ibc inc & Type Ib & Type Ic & Unresolved Type Ibc & Type Ic-bl & All core-collapse\\
\hline
Type IIP\\
Type IIb &     0.0677\\
Type IIn &     0.0302 &     0.6571\\
Type IIb \& IIn &     0.0204 &     0.9193 &     1.0000\\
Type Ibc inc &     0.9061 &     0.0686 &     0.1827 &     0.1269\\
Type Ib &     0.7984 &     0.1638 &     0.1994 &     0.1801 &     0.9857\\
Type Ic &     0.9407 &     0.1088 &     0.4814 &     0.2806 &     1.0000 &     0.8204\\
Unresolved Type Ibc &     0.9895 &     0.1668 &     0.4140 &     0.2655 &     0.9894 &     0.9678 &     0.9804\\
Type Ic-bl &     0.9987 &     0.7014 &     0.5450 &     0.5654 &     0.9979 &     0.9597 &     0.9869 &     0.9986\\
All core-collapse &     0.6810 &     0.0685 &     0.0645 &     0.0258 &     0.9920 &     0.8748 &     0.9898 &     0.9233 &     0.9914\\
All Type I &     0.1236 &     0.1272 &     0.4404 &     0.1337 &     0.6002 &     0.5580 &     0.9144 &     0.7726 &     0.8933 &     0.0425\\

\hline
\end{tabular}
%\end{center}
\vspace{-0.2 cm}
%\hspace{-3.5 cm}
\hspace{-0.7 cm}
\hspace{1.48 cm}\begin{minipage}[H]{0.9\textwidth}
{\vspace{0.4 cm}\normalsize Computed Kolmogorov--Smirnov (KS) probabilities comparing the metallicity distributions of various SNe Types.  Values of ``1.0000" are the result of rounding.  ``Inc" includes subtypes and ``Exc" does not (i.e.\ a SN Type IIP would be counted as a `Type II inc" but not as a `Type II exc").}
\end{minipage}

\end{sidewaystable*}

\cleardoublepage

%\onecolumn

%\newcolumntype{L}[1]{>{\centering\arraybackslash}m{#1}}
\newcolumntype{S}[1]{@{\hskip 0.3 cm}m{#1}}
\begin{longtable*}[p]{m{1.9cm}L{1.7cm}L{1.7cm}L{1.7cm}L{1.7cm}L{1.7cm}m{1.9cm}}
%\begin{center}
%\vspace{-0.2 cm}
%\begin{minipage}[H]{0.45\textwidth}
\caption{\\ \label{SNe_Z_table} List of supernovae with metallicities  \vspace{-0.23 cm}}
%\end{minipage}
%\begin{tabular}{m{3.3cm}L{1.7cm}L{1.7cm}}
%\cline{1-2}
\\\hline
\hline
\vspace{0.002 cm}
\multirow{2.7}{*}{object} & redshift & M$_B$ & \multicolumn{3}{c}{12+log(O/H)} & \multirow{2.3}{*}{SNe Type} \\
\cline{4-6}
\vspace{0.002 cm}
& (z) & (mag) & \citetalias{KobulnickyKewley} & \citetalias{T04} & \citetalias{Dopita2016} &  \\
\hline
\endhead
%SN 2200asv & 0.0934 & 8.4 & 9.6 & 7.8 &  Ia bhtlv like daft \\
%\input{SNe_Z_table.txt}
SN 1959C &     0.0101 &     -17.43 &      &       8.48 &       8.12 & Ia \\
SN 1961F &     0.0047 &     -17.72 &      &       8.67 &       8.29 & II \\
SN 1964L &     0.0026 &     -14.04 &      &       9.19 &       8.87 & Ic \\
SN 1982W &     0.2176 &     -21.78 &       8.91 &       8.81 &       8.55 & I \\
SN 1984A &    -0.0002 &      -0.16 &      &     &       8.79 & I \\
SN 1985B &     0.0065 &     -19.09 &      &     &       9.21 & II \\
SN 1985F &     0.0018 &     -16.88 &      &       8.73 &       8.41 & Ib \\
SN 1985G &     0.0026 &     -11.72 &      &       9.05 &       8.64 & II \\
\end{longtable*}

\newcolumntype{L}[1]{>{\centering\arraybackslash}m{#1}}
\newcolumntype{J}[1]{>{\centering\arraybackslash}m{#1}}
%\newcolumntype{J}[1]{m{#1}\arraybackslash}
\begin{longtable*}[p]{p{1.9cm}L{1.35cm}L{1.2cm}L{1.2cm}L{1.25cm}L{1.2cm}L{1.25cm}L{1.2cm}L{1.2cm}m{1.15cm}}
\caption{\\ \label{SNe_line_table} List of spectral line fluxes for supernovae with metallicities in \aref{SNe_Z_table} \vspace{-0.23 cm}}
%\cline{1-3}
\\\hline
\hline
\vspace{0.02 cm}
\multirow{2.3}{*}{object} & [O~II] & H$\beta$ & \multicolumn{2}{c}{[O~III]} & [O~I] & H$\alpha$ & [N~II] & \multicolumn{2}{c}{[S~II]} \\
\cline{4-5}\cline{9-10}
\vspace{0.002 cm}
& 3727 {\AA} & 4861 {\AA} & 4959  {\AA} & 5007  {\AA} & 6300  {\AA} &  6563  {\AA}& 6584  {\AA} & 6717  {\AA} & 6731  {\AA} \\
\hline
\endhead
%SN 2200asv & 5.43 & 0.032 & 9.543 & 5.43 & 0.032 & 9.543 & 5.43 & 0.032 \\
%\input{SNe_line_table.txt}
%\input{SNe_line_table.txt}
%\input{SNe_line_table_dumb.txt}
\input{SNe_line_table_w_errors.txt}
\hline
\end{longtable*}
\begin{minipage}[H]{0.45\textwidth}
\vspace{0.1 cm}
\phantomsection\label{SNe_line_table_end}
\end{minipage}

\label{LastPage}
\end{document}